\documentclass[12pt]{JHEP3}
\usepackage{amssymb,amsmath,amsthm}
\usepackage{mathrsfs} 
\def\TODAY{1 September 2010}

\title{\bf Generic master equations for quasi-normal frequencies}
\author{{\bf Jozef Skakala} and {\bf Matt Visser}\\
School of Mathematics, Statistics, and Operations Research, \\
Victoria University of Wellington, \\
Wellington, New Zealand\\
E-mail: \email{jozef.skakala@msor.vuw.ac.nz, matt.visser@msor.vuw.ac.nz}
}
\date{\TODAY;  \LaTeX-ed \today}                                           
\maketitle
\abstract{
Generic master equations governing the highly-damped quasi-normal frequencies [QNFs] of one-horizon, two-horizon, and even three-horizon spacetimes can be obtained through either semi-analytic or monodromy techniques. While many technical details differ, both between the semi-analytic and monodromy approaches,  and quite often among various authors seeking to apply the monodromy technique, there is nevertheless widespread agreement regarding the general form of the QNF master equations.  Within this class of generic master equations we can establish some rather general results, relating the existence of ``families'' of QNFs of the form
\[
\omega_{a,n} =  \hbox{(offset)}_a + i n \; \hbox{(gap)}
\]
to the question of whether or not certain ratios of parameters are rational or irrational.
}


\begin{document}
\clearpage
\def\d{{\mathrm{d}}}
\newcommand{\scri}{\mathscr{I}}
\newcommand{\sun}{\ensuremath{\odot}}
\def\J{{\mathscr{J}}}
\def\sech{{\mathrm{sech}}}
\def\T{{\mathcal{T}}}
\tableofcontents
\clearpage
\section{Introduction}
Generic master equations governing the highly-damped quasi-normal frequencies [QNFs] of one-horizon, two-horizon, and even three-horizon spacetimes can be obtained through either semi-analytic~\cite{Skakala1, Skakala2, Suneeta} or monodromy techniques~\cite{Motl1, Motl2, Musiri, Andersson, Choudhury, Cardoso, NS, KR, Das, Lopez-Ortega, Shu-Shen, Green, review}. (For general background, see also~\cite{Shanka1, Shanka2, Chan1, Chan2, Horowitz, Wang1, Wang2, Wang3, Wang4, Wang5, Wang6}.) In these approaches one either approximates the true gravitational potential by a piecewise exactly solvable potential, or one works with an analytic continuation into the complex radial plane. While many technical details differ, both between the semi-analytic and monodromy approaches,  and often among various authors seeking to apply the monodromy technique, there is widespread agreement that both approaches lead to QNF master equations of the general form:
\begin{equation}
\sum_{A=1}^N   C_A \; \exp\left(  \sum_{i=1}^H  {Z_{Ai} \; \pi \omega\over \kappa_i} \right) = 0.
\label{E:master}
\end{equation}
Here $\kappa_i$ is the surface gravity of the $i$'th horizon, $H$ is the number of horizons, the matrix $Z_{Ai}$ always has rational entries (and quite often is integer-valued). The physics contained in the master equation is invariant under substitutions of the form $Z_{Ai}\to Z_{Ai} + (1, \dots, 1)_A^T \; h_i$, where the $h_i$ are arbitrary rational numbers. Either $\sum_A Z_{Ai}=0$, or it can without loss of generality be made zero.  Furthermore $N$ is some reasonably small positive integer. (In fact $N\leq2H+1$ in all situations we have encountered, and typically $N>H$.) The $C_A$ are a collection of coefficients that are often but not always integers, though in all known cases they are at least real. Finally in almost all known cases the rectangular $N\times H$ matrix $Z_{Ai}$ has rank $H$, and the QNF master equation is almost always irreducible (that is, non-factorizable). 

We shall initially demonstrate that all known master equations (whether based on semi-analytic or monodromy techniques) can be cast into this form. We shall then explore what can be said concerning the solutions of this generic class of QNF conditions.  Specifically, if the ratios $R_{ij} = \kappa_i/\kappa_j$ are all rational numbers, then the master equation can be converted into a polynomial and the QNFs will automatically fall into ``families'' of the form
\begin{equation}
\omega_{a,n} =  \hbox{(offset)}_a + i n \; \hbox{(gap)};  \qquad n\in\{0,1,2,,3,\dots\}
\label{E:family}
\end{equation}
Conversely, if even one such family exists, this places very strong constraints on the ratios $R_{ij} = \kappa_i/\kappa_j$.

\section{Generic master equations for QNFs}\label{S:generic}
\subsection{Semi-analytic techniques}\label{SS:semi-analytic}

Semi-analytic techniques approximate the real physical problem by breaking it down into segments that are piecewise exactly solvable~\cite{Skakala1, Skakala2, Suneeta}. 

\paragraph{One horizon:}
In a one-horizon situation the effective gravitational potential has a exponential falloff in one direction and an inverse-square falloff in the other direction. 
The effective gravitational potential is approximated by
\begin{equation}
V(x) = \left\{  \begin{array}{lcl}
{V_{0-} \; \sech^2(\kappa x)}  & \hbox{ for }  & x < 0;\\
V_{0+} \;  {a^2\over(x+a)^2} & \hbox{ for }  & x > 0. \\
\end{array}\right.
\end{equation}
For highly damped QNFs the master equation is derived in reference~\cite{Skakala2}, in a form equivalent to
\begin{equation}
\sinh\left({\pi\omega\over\kappa}\right)=0.
\end{equation}

\paragraph{Two horizons:}
For the two-horizon case the potential is approximated by
\begin{equation}
V(x) = \left\{  \begin{array}{lcl}
{V_{0-} \; \sech^2(\kappa_-\, x)}  & \hbox{ for }  & x < 0;\\
V_{0+} \; \sech^2(\kappa_+\, x)  & \hbox{ for }  & x > 0. \\
\end{array}\right.
\end{equation}
Both $\kappa_\pm$ are by convention positive.
For highly damped QNFs the master equation is derived in references~\cite{Skakala1, Skakala2} in a form equivalent to
\begin{equation}
\cosh\left({\pi\omega\over\kappa_+} + {\pi\omega\over\kappa_-}\right) - \cosh\left({\pi\omega\over\kappa_+}- {\pi\omega\over\kappa_-}\right) 
+ 2  \;  \cos(\pi\alpha_+)\cos(\pi\alpha_-) = 0,
\end{equation}
where 
\begin{equation}
\alpha_\pm = \sqrt{{1\over4} - {V_{0\pm}\over\kappa_\pm^2}}; \qquad \alpha_\pm \neq {1\over2}. 
\end{equation}
Note that $\alpha_\pm = {1\over2}$ is a physically degenerate case corresponding either to $V_{0\pm}=0$ (in which case the corresponding $\kappa_\pm$ is physically and mathematically meaningless), or $\kappa_\pm = \infty$, (in which case the QNF master equation is vacuous).  Either of these situations is unphysical so one must have  $\alpha_\pm \neq {1\over2}$.  This QNF condition above is irreducible (non-factorizable) unless $\alpha_\pm = m+{1\over2}$ with $m\in Z$. This occurs when
\begin{equation}
V_{0\pm} = - m(m+1) \; \kappa_\pm^2,
\end{equation}
and in this exceptional situation the QNF master equation factorizes to
\begin{equation}
\sinh\left({\pi\omega\over\kappa_+} \right) \sinh\left( {\pi\omega\over\kappa_-}\right) = 0. 
\end{equation}
In this case the QNF spectrum becomes
\begin{equation}
\label{E:2s}
\omega_{n_+,n_-} = \left\{ \begin{array}{c}  i  n_+ \kappa_+; \\  i n_- \kappa_-; \end{array}\right.
\end{equation}
with no restriction on the relative values of $\kappa_\pm$. 
This factorizable  situation is however clearly non-generic in the semi-analytic approach.

\paragraph{Three horizons:} There is no simple or practicable way of dealing with three--horizon situations suing semi-analytic techniques. 

\paragraph{Summary:}
In both applicable cases the hyperbolic functions can be traded for exponentials and the QNF master condition can be put in the form of equation (\ref{E:master}), with either $H=1$ and $N=2$ terms,  or $H=2$ and $N=5$ terms respectively. The vectors $C_A$ and matrices $Z_{Ai}$ are
\begin{equation}
C_{A}= \left[
\begin{array}{c}
+1  \\ -1 
\end{array}
\right]
\qquad
Z_{A1}= \left[
\begin{array}{c}
+1  \\ -1 
\end{array}
\right],
\end{equation}
and
\begin{equation}
C_{A}= \left[
\begin{array}{c}
+1  \\
-1 \\
2  \;  \cos(\pi\alpha_+)\cos(\pi\alpha_-)\\
-1 \\
+1\\
\end{array}
\right]
\qquad 
Z_{Ai}= \left[
\begin{array}{rr}
+1  & +1 \\
+1 & -1\\
0 & 0 \\
-1 & +1\\
-1 & -1 \\
\end{array}
\right],
\end{equation}
respectively. (Both matrices $Z_{Ai}$ have maximal rank $H$, and both satisfy $\sum_A Z_{Ai}=0$.)
In almost all physically relevant situations ($\alpha_\pm\neq m+ {1\over2}$ with $m\in Z$) this QNF condition is irreducible (non-factorizable).

\subsection{Monodromy techniques}\label{SS:monodromy}

Monodromy techniques involve an analytic continuation into the complex plane, together with information regarding the singular points of the relevant ordinary differential equation~\cite{Motl1, Motl2, Musiri, Andersson, Choudhury, Cardoso, NS, KR, Das, Lopez-Ortega, Shu-Shen, Green, review}. 

\paragraph{One horizon:}
In the one-horizon situation there is general agreement that the relevant master equation is
\begin{equation}
\exp\left({\pi\omega\over\kappa}\right) + 1 + 2\cos(\pi j) = 0.
\end{equation}
Unfortunately there is distressingly little agreement over the precise status of the parameter $j$. References~\cite{Motl1, Motl2, Musiri,  Andersson, NS, review} assert that this is the spin of the perturbation under consideration, but with some disagreement as to whether this applies to all spins and all dimensions.  

Note that the phrase ``perturbation under consideration'' might refer to perturbations of intrinsically linear situations around trivial backgrounds (e.g., spin zero scalars satisfying the Klein--Gordon equations, spin 1 vectors satisfying the Maxwell equations), or to linearizations of intrinsically nonlinear problems (e.g., spin 2  perturbations of the spacetime geometry described by the Regge--Wheeler or Zerelli equations).  Most of the relevant literature confines itself to perturbations of spin $\{0,1,2\}$, but there is no intrinsic obstruction to considering higher-spin situations.

In contrast in reference~\cite{Das} a particular model for the spacetime metric is adopted, and in terms of the parameters describing this model these authors take
\begin{equation}
j = {qd\over2}-1.
\end{equation}
Reference~\cite{Lopez-Ortega} asserts that for spin 1 perturbations
\begin{equation}
j = {2(d-3)\over d-2}.
\end{equation}
Be this as it may, there is universal agreement on the \emph{form} of the QNF master condition, and it  is automatically of the the form of equation (\ref{E:master}), with $H=1$ and $N=2$ terms. The vector $C_A$ and the matrix $Z_{Ai}$ are
\begin{equation}
C_{A}= \left[
\begin{array}{c}
+1  \\ 1+2\cos(\pi j)
\end{array}
\right];
\qquad
Z_{A1}= \left[
\begin{array}{r}
+1  \\ 0 
\end{array}
\right].
\end{equation}
By multiplying through by $\exp(-\pi\omega/(2\kappa))$ we can re-cast the QNF condition as
\begin{equation}
\exp\left({\pi\omega\over2\kappa}\right) + \{1 + 2\cos(\pi j)\} \exp\left(-{\pi\omega\over2\kappa}\right)  = 0.
\end{equation}
This now corresponds to 
\begin{equation}
C_{A}= \left[
\begin{array}{c}
+1  \\ 1+2\cos(\pi j)
\end{array}
\right];
\qquad
Z_{A1}= \left[
\begin{array}{r}
+1/2  \\ -1/2 
\end{array}
\right],
\end{equation}
and in this form we have $\sum_A Z_{Ai}=0$. (This is one of rather few cases where it is convenient to take the $Z_{Ai}$ to be rational-valued rather than integer-valued.)

\paragraph{Two horizons:}
For two-horizon situations the analysis is slightly different for Schwarzschild--de~Sitter spacetimes (Kottler spacetimes) versus Reissner--Nordstr\"om spacetimes. 
\begin{itemize}
\item 
For {\bf Schwarzschild--de~Sitter} spacetimes there is general agreement that the relevant master equation for the QNFs is
\begin{equation}
\label{E:SdS}
\{1 + 2\cos(\pi j) \}\cosh\left({\pi\omega\over\kappa_+} + {\pi\omega\over\kappa_-}\right) + \cosh\left({\pi\omega\over\kappa_+}- {\pi\omega\over\kappa_-}\right) 
=  0.
\end{equation}
We shall again adopt conventions such that $\kappa_\pm$ are both positive. 
Again, there is unfortunately distressingly little agreement over the precise status of the parameter $j$. References~\cite{Motl1, Motl2, Musiri, Andersson, NS, review} assert that this is the spin of the perturbation under consideration, but with some disagreement as to whether this applies to all spins and all dimensions.  In contrast in reference~\cite{Das} a particular model for the spacetime metric is again adopted, and in terms of the parameters describing this model they take
\begin{equation}
j = {qd\over2}-1.
\end{equation}
One still has to perform a number of trigonometric transformations to turn the quoted result of reference~\cite{Das} for $d\neq 5$
\begin{equation}
\tanh\left({\pi\omega\over\kappa_+}\right) \tanh\left( {\pi\omega\over\kappa_-}\right) = {2\over\tan^2(\pi j/2)-1},
\end{equation}
into the equivalent form (\ref{E:SdS}) above. 
For $d=5$ the authors of~\cite{Shu-Shen} assert the equivalent of
\begin{equation}
\label{E:SdS5}
\{1 + 2\cos(\pi j) \}\sinh\left({\pi\omega\over\kappa_+} + {\pi\omega\over\kappa_-}\right) + \sinh\left({\pi\omega\over\kappa_+}- {\pi\omega\over\kappa_-}\right) 
=  0.
\end{equation}
Reference~\cite{Lopez-Ortega} again asserts that for spin 1 perturbations
\begin{equation}
j = {2(d-3)\over d-2}.
\end{equation}
Be this as it may, there is again universal agreement on the \emph{form} of the QNF master condition, and converting hyperbolic functions into exponentials, it can be transformed into  the form of equation (\ref{E:master}), with $H=2$ and $N=4$ terms.
Focusing on the generic case of equation (\ref{E:SdS}), the vector $C_A$ and matrix $Z_{Ai}$ are
\begin{equation}
C_{A}= \left[
\begin{array}{c}
1+2\cos(\pi j) \\
+1 \\
+1\\
1+2\cos(\pi j) \\
\end{array}
\right];
\qquad
Z_{Ai}= \left[
\begin{array}{rr}
+1  & +1 \\
+1 & -1\\
-1 & +1\\
-1 & -1 \\
\end{array}
\right].
\end{equation}
Note that we explicitly have $Z_{Ai}=0$. 
There are two exceptional cases:
\begin{itemize}
\item 
If $j = 2m+1$ with $m\in Z$ then 
\begin{equation}
C_{A}= \left[
\begin{array}{c}
-1\\
+1 \\
+1\\
-1\\
\end{array}
\right];
\qquad
Z_{Ai}= \left[
\begin{array}{rr}
+1  & +1 \\
+1 & -1\\
-1 & +1\\
-1 & -1 \\
\end{array}
\right].
\end{equation}
In this situation the QNF master equation factorizes
\begin{equation}
\sinh\left({\pi\omega\over\kappa_+} \right) \sinh\left( {\pi\omega\over\kappa_-}\right) = 0. 
\end{equation}
This appears to be the physically relevant case for spin 1 perturbations. The relevant QNF spectrum is that of equation (\ref{E:2s}). 

\item
If $\cos(\pi j) = - {1\over2}$, which does not appear to be a physically relevant situation but serves to illustrate potential mathematical pathologies, then
\begin{equation}
C_{A}= \left[
\begin{array}{r}
0\\
+1 \\
+1\\
0\\
\end{array}
\right];
\qquad
Z_{Ai}= \left[
\begin{array}{rr}
+1  & +1 \\
+1 & -1\\
-1 & +1\\
-1 & -1 \\
\end{array}
\right].
\end{equation}
But in this situation the top row and bottom row do not contribute to the QNF master equation and one might as well delete them. That is, one might as well write
\begin{equation}
C_{A}= \left[
\begin{array}{r}
+1 \\
+1\\
\end{array}
\right];
\qquad
Z_{Ai}= \left[
\begin{array}{rr}
+1 & -1\\
-1 & +1\\
\end{array}
\right].
\end{equation}
This is a situation (albeit unphysical) where the matrix $Z_{Ai}$ does not have maximal rank. The QNF master equation degenerates to
\begin{equation}
\sinh\left({\pi\omega\over\kappa_+}- {\pi\omega\over\kappa_-}\right) 
=  0.
\end{equation}
In this situation the QNF spectrum is
\begin{equation}
\omega_n = { i n \kappa_+ \kappa_- \over |\kappa_+ - \kappa_-|},
\end{equation}
with no restriction on the relative values of $\kappa_\pm$. 
This situation is however clearly non-generic (and outright unphysical).

\end{itemize}

\item
For {\bf Reissner--Nordstr\"om} spacetime one has~\cite{Motl2}
\begin{equation}
\label{E:RN}
\exp\left({2\pi\omega\over\kappa_+}\right) 
+2 \{1 + \cos(\pi j) \}\exp\left(- {2\pi\omega\over\kappa_-}\right) + \{1 + 2\cos(\pi j) \}=  0,
\end{equation}
where $\kappa_+$ is the surface gravity of the outer horizon and $\kappa_-$ is the surface gravity of the inner horizon. 
There is again some disagreement on the status of the parameter $j$. For the perturbations under consideration reference~\cite{Motl2} now takes $j={1\over3}$ for spin 0, and $j={5\over3}$ for spins 1 and 2 (in any dimension). 
Reference~\cite{NS} asserts that for general dimension 
\begin{equation}
j=\frac{d-3}{2d-5} \qquad \hbox{for spin 0, 2, and}
\qquad
j=\frac{3d-7}{2d-5}\qquad \hbox{ for spin 1}.
\end{equation}
Be this as it may, there is universal agreement on the \emph{form} of the QNF master condition, and it  is automatically of the form of equation (\ref{E:master}), with $H=2$ and $N=3$ terms.
The vector $C_A$ and matrix $Z_{Ai}$ are
\begin{equation}
C_{A}= \left[
\begin{array}{c}
+1   \\
2\{1+\cos(\pi j)\} \\
1+2\cos(\pi j)\\
\end{array}
\right];
\qquad
Z_{Ai}= \left[
\begin{array}{rr}
+2  & 0 \\
0 & -2\\
0 & 0\\
\end{array}
\right].
\end{equation}
If we multiply through by a suitable factor then we can write the QNF condition in the equivalent form
\begin{eqnarray}
\label{E:RN3}
&&\exp\left({2\pi\omega\over3\kappa_+} + {\pi\omega\over3\kappa_-}\right) 
+2 \{1 + \cos(\pi j) \}\exp\left(  - {\pi\omega\over3\kappa_+}- {2\pi\omega\over3\kappa_-}\right) 
\nonumber
\\
&&
\qquad + \{1 + 2\cos(\pi j) \}  \exp\left( - {\pi\omega\over3\kappa_+} + {\pi\omega\over3\kappa_-} \right)=  0,
\end{eqnarray}
This corresponds to 
\begin{equation}
C_{A}= \left[
\begin{array}{c}
+1   \\
2\{1+\cos(\pi j)\} \\
1+2\cos(\pi j)\\
\end{array}
\right];
\qquad
Z_{Ai}= \left[
\begin{array}{rr}
+2/3  & 1/3 \\
-1/3 & -2/3\\
-1/3& 1/3\\
\end{array}
\right].
\end{equation}
In this form we now explicitly have $\sum_A Z_{Ai}=0$. (This is one of rather few cases where it is convenient to take the $Z_{Ai}$ to be rational-valued rather than integer-valued.)
Returning to the original form in equation (\ref{E:RN}), there are two exceptional cases:
\begin{itemize}
\item 
If $j = 2m+1$ with $m\in Z$, (this does not appear to be a physically relevant situation but again this serves to illustrate the possible mathematical pathologies one might encounter), then 
\begin{equation}
C_{A}= \left[
\begin{array}{r}
+1\\
0 \\
-1\\
\end{array}
\right];
\qquad
Z_{Ai}= \left[
\begin{array}{rr}
+2  & 0 \\
0 & -2\\
0 & 0\\
\end{array}
\right].
\end{equation}
But then (without loss of information) one might as well eliminate the second row, to obtain
\begin{equation}
C_{A}= \left[
\begin{array}{r}
+1\\
-1\\
\end{array}
\right];
\qquad
Z_{Ai}= \left[
\begin{array}{rr}
+2  & 0 \\
0 & 0\\
\end{array}
\right].
\end{equation}
Furthermore, since $\kappa_-$ now decouples,  we might as well eliminate the second column, to obtain
\begin{equation}
C_{A}= \left[
\begin{array}{r}
+1\\
-1\\
\end{array}
\right];
\qquad
Z_{Ai}= \left[
\begin{array}{rr}
2  \\
0 \\
\end{array}
\right].
\end{equation}
The QNF master equation then specializes to
\begin{equation}
\label{E:RN-sp1}
\exp\left({2\pi\omega\over\kappa_+}\right) -1 =  0.
\end{equation}

\item
If $\cos(\pi j) = - {1\over2}$, which does not appear to be a physically relevant situation but serves to illustrate potential mathematical pathologies, then
\begin{equation}
C_{A}= \left[
\begin{array}{r}
+1 \\
+1\\
0\\
\end{array}
\right];
\qquad
\qquad
Z_{Ai}= \left[
\begin{array}{rr}
+2  & 0 \\
0 & -2\\
0 & 0\\
\end{array}
\right].
\end{equation}
But in this situation the  bottom row does not contribute to the QNF master equation and one might as well delete it. That is, one might as well write
\begin{equation}
C_{A}= \left[
\begin{array}{r}
+1 \\
+1\\
\end{array}
\right];
\qquad
\qquad
Z_{Ai}= \left[
\begin{array}{rr}
+2  & 0 \\
0 & -2\\
\end{array}
\right].
\end{equation}
We can rearrange the terms in the master equation to have 
the QNF master equation specialize to 
\begin{equation}
\label{E:RN-sp2}
\exp\left({2\pi\omega\over\kappa_+} + {2\pi\omega\over\kappa_-}\right) +1 =  0.
\end{equation}
This corresponds to
\begin{equation}
C_{A}= \left[
\begin{array}{r}
+1 \\
+1\\
\end{array}
\right];
\qquad
\qquad
Z_{Ai}= \left[
\begin{array}{rr}
+2  & +2 \\
0 & 0 \\
\end{array}
\right].
\end{equation}
Note that in this exceptional case $Z_{Ai}$ is not of maximal rank.
In this situation the QNF spectrum is
\begin{equation}
\omega_n = {(2 n+1)i \;\kappa_+ \kappa_- \over \kappa_+ + \kappa_-},
\end{equation}
with no restriction on the relative values of $\kappa_\pm$. 
This situation is however clearly non-generic (and outright unphysical).

\end{itemize}

\end{itemize}

\paragraph{Three horizons:}
For three horizons the natural example to consider is that of  Reissner--Nordstr\"om--de~Sitter spacetime. References~\cite{NS, Shu-Shen} agree that (for $d\neq 5$)
\begin{eqnarray}
\label{E:RNdS}
&&\cosh\left({\pi\omega\over\kappa_+}- {\pi\omega\over\kappa_0}\right) 
+ \{1 + \cos(\pi j) \} \cosh\left({\pi\omega\over\kappa_+}+ {\pi\omega\over\kappa_0}\right)  
\nonumber\\
&& \qquad\qquad + 2 \{1 + \cos(\pi j) \}  \cosh\left({2\pi\omega\over\kappa_-}+{\pi\omega\over\kappa_+}+ {\pi\omega\over\kappa_0}\right) =  0.
\end{eqnarray}
Here $\kappa_\pm$ refer to the inner and outer horizons of the central Riessner--Nordstr\"om black hole, while $\kappa_0$ is now the surface gravity of the cosmological horizon. All these surface gravities are taken positive.  In contrast for $d=5$ one has
\begin{eqnarray}
\label{E:RNdS5}
&&\sinh\left({\pi\omega\over\kappa_+}- {\pi\omega\over\kappa_0}\right) 
+ \{1 + \cos(\pi j) \} \sinh\left({\pi\omega\over\kappa_+}+ {\pi\omega\over\kappa_0}\right)  
\nonumber\\
&& \qquad\qquad + 2 \{1 + \cos(\pi j) \}  \sinh\left({2\pi\omega\over\kappa_-}+{\pi\omega\over\kappa_+}+ {\pi\omega\over\kappa_0}\right) =  0,
\end{eqnarray}
Again, for the various perturbations under consideration
\begin{equation}
j=\frac{d-3}{2d-5} \qquad \hbox{for spin 0, 2, and}
\qquad
j=\frac{3d-7}{2d-5}\qquad \hbox{ for spin 1}.
\end{equation}
There is universal agreement on the \emph{form} of the QNF master condition, and converting hyperbolic functions into exponentials, it can be transformed into  the form of equation (\ref{E:master}), with $H=3$ and $N=6$ terms.
The vector $C_A$ and matrix $Z_{Ai}$ are
\begin{equation}
C_{A}= \left[
\begin{array}{c}
+1\\
1+\cos(\pi j)  \\
2\{1+\cos(\pi j)\} \\
\pm 2\{1+\cos(\pi j)\} \\
\pm\{1+\cos(\pi j) \}\\
\pm 1\\
\end{array}
\right];
\qquad
Z_{Ai}= \left[
\begin{array}{rrr}
+1  & -1 & 0 \\
+1 & +1 & 0\\
+1 & +1 & +2\\
-1 & -1 & -2\\
-1 & -1 & 0\\
-1 & +1 & 0\\
\end{array}
\right]. 
\end{equation}
Generically, $Z_{Ai}$ has maximal rank $H=3$. Note that we explicitly have $\sum_A Z_{Ai}=0$. 

The only exceptional case is $\cos(\pi j)= -1$ in which case
\begin{equation}
C_{A}= \left[
\begin{array}{c}
+1\\
0  \\
0\\
0 \\
0 \\
\pm 1\\
\end{array}
\right];
\qquad
Z_{Ai}= \left[
\begin{array}{rrr}
+1  & -1 & 0 \\
+1 & +1 & 0\\
+1 & +1 & +2\\
-1 & -1 & -2\\
-1 & -1 & 0\\
-1 & +1 & 0\\
\end{array}
\right]. 
\end{equation}
But then the $2^{nd}$ to $5^{th}$ rows decouple and may as well be removed, yielding
\begin{equation}
C_{A}= \left[
\begin{array}{c}
+1\\
\pm 1\\
\end{array}
\right];
\qquad
Z_{Ai}= \left[
\begin{array}{rrr}
+1  & -1 & 0 \\
-1 & +1 & 0\\
\end{array}
\right]. 
\end{equation}
The $3^{rd}$ column, corresponding to $\kappa_-$, now decouples and may as well be removed, yielding
\begin{equation}
C_{A}= \left[
\begin{array}{c}
+1\\
\pm 1\\
\end{array}
\right];
\qquad
Z_{Ai}= \left[
\begin{array}{rr}
+1  & -1  \\
-1 & +1 \\
\end{array}
\right]. 
\end{equation}
Note that in this exceptional case $Z_{Ai}$ is not of maximal rank, and the QNF master equation degenerates to
\begin{equation}
\cosh\left({\pi\omega\over\kappa_+}- {\pi\omega\over\kappa_0}\right) 
= 0,
\qquad
\hbox{or}
\qquad
\sinh\left({\pi\omega\over\kappa_+}- {\pi\omega\over\kappa_0}\right) 
= 0,
\end{equation}
respectively.
In this situation the QNF spectrum is
\begin{equation}
\omega_n = {(2n+1) i \; \kappa_+ \kappa_0 \over 2 |\kappa_+ - \kappa_0|},
\quad
\hbox{or}
\qquad
\omega_n = { i n \kappa_+ \kappa_0 \over |\kappa_+ - \kappa_0|},
\end{equation}
respectively, 
with no restriction on the relative values of $\kappa_\pm$. 
This situation is however clearly non-generic (and outright unphysical).

\subsection{Summary: General form}\label{SS:general}

In summary, every example we have seen can be cast in the form
\begin{equation}
\sum_{A=1}^N   C_A \; \exp\left(  \sum_{i=1}^H  {Z_{Ai} \; \pi \omega\over \kappa_i} \right) = 0,
\label{E:master2}
\end{equation}
where $N\leq 2H+1$ in all the situations we have encountered. Typically $N>H$. Remember, $\kappa_i$ is the surface gravity of the $i$'th horizon, $H$ is the number of horizons, the $Z_{Ai}$ are integers (or at worst rational), and the $C_A$ are a collection of coefficients that are often but not always integers, though in all known cases they are at least real. The matrix $Z_{Ai}$ is generically of maximal rank $H$ in all situations we have encountered --- physically this seems to be due to the fact that if the matrix $Z_{Ai}$ is \emph{not} of maximal rank, then this implies that some linear combination of the (inverse) surface gravities completely decouples from the QNF master equation --- which is arguably unphysical. 
By multiplying the equation through by a common factor of $\exp( \sum_{i=1}^H  {h_i \; \pi \omega/ \kappa_i}  )$  it becomes clear that the physics encoded in the master equation remains  invariant under substitutions of the form $Z_{Ai}\to Z_{Ai} + (1,\dots,1)_A^T \; h_i$.  By choosing the $h_i$ to be the rational numbers $h_i = - (\sum_A Z_{Ai})/N$ we can always arrange for the shifted $Z_{Ai}$ to satisfy $\sum_A Z_{Ai} \to 0$.
Furthermore in all situations encountered to date this master equation is generically irreducible (non-factorizable). 
It is this general form of the master equation that we shall now analyze in detail to place as  many constraints as possible on the QNFs.

\def\m{{\tilde m}}

\section{From master equation to polynomial}\label{S:polynomial}

First let us suppose that the ratios $R_{ij} = \kappa_i/\kappa_j$ are all rational numbers. This is not as significant a constraint as one might initially think. In particular, since the rationals are dense in the reals one can always with arbitrarily high accuracy make an approximation to this effect.  Furthermore since floating point numbers are essentially a subset of the rationals, all numerical investigations implicitly make such an assumption, and all numerical experiments should be interpreted with this point kept firmly in mind.

Provided that the ratios $R_{ij} = \kappa_i/\kappa_j$ are all rational numbers, it follows that there is a constant $\kappa_*$ and a collection of relatively prime integers $m_i$ such that
\begin{equation}
\kappa_i = {\kappa_*\over m_i}.
\end{equation}
The QNF master equation then becomes
\begin{equation}
\sum_{A=1}^N   C_A \; \exp\left(  \sum_{i=1}^H  Z_{Ai}  \; m_i \; {\pi \omega\over \kappa_*} \right) = 0,
\label{E:master2-1}
\end{equation}
Now define $z=\exp(\pi\omega/\kappa_*)$,  and define a new set of integers $\m_A  =  \sum_{i=1}^H  Z_{Ai}  \; m_i $. (There is no guarantee or requirement that the  $\m_A $ be relatively prime, and some of the special cases we had to consider in reference~\cite{Skakala1} ultimately depend on this observation.)
Then
\begin{equation}
\sum_{A=1}^N   C_A \; z^{\m_A} = 0.
\label{E:master2-2}
\end{equation}
This is (at present) a Laurent polynomial, as some exponents may be (and typically are) negative. Multiplying through by $z^{-\tilde m_\mathrm{min}}$ converts this to a regular polynomial with a nonzero constant $z^0$ term and with degree
\begin{equation}
D = \tilde m_\mathrm{max} - \tilde m_\mathrm{min}. 
\end{equation}
If we write $\bar m_A = \m_A - \tilde m_\mathrm{min}$ then the relevant regular polynomial is
\begin{equation}
\sum_{A=1}^N   C_A \; z^{\bar m_A} = 0.
\label{E:master2-3}
\end{equation}
Note that the polynomial is typically ``sparse'' --- the number of terms $N$ is small (typically $N\leq 2H+1$) but the degree $D$ can easily be arbitrarily large. 
There are at most $D$ distinct roots for the polynomial $z_a$, and the \emph{general} solution of the QNF condition is
\begin{equation}
\omega_{a,n} = {\kappa_*\ln(z_a)\over\pi} + {2in \kappa_*}; \qquad a\in\{1,...,D\}; \qquad n\in\{0,1,2,3,\dots\}.
\end{equation}
If the $\bar m_A$ are not relatively prime, define a degeneracy factor $g= \mathrm{hcf}\{\bar m_A\}$. Then the roots will fall into $D/g$ classes where the $g$ degenerate members of each class differ only by the various $g$-th roots of unity. In this situation we can somewhat simplify the above QNF spectrum to yield
\begin{equation}
\omega_{a,n} = {\kappa_*\ln(z_a)\over\pi} + {2in \kappa_*\over g}; \qquad a\in\{1,...,D/g\}; \qquad n\in\{0,1,2,3,\dots\}.
\end{equation}
We again emphasize that behaviour of this sort certainly does occur in practice. There is no guarantee or requirement that the  $\bar m_A $ be relatively prime, and some of the special cases we had to consider in reference~\cite{Skakala1} ultimately depend on this observation.

There is a (slightly) weaker condition that also leads to polynomial master equations and the associated families of QNFs. 
Suppose that we know that the ratios 
\begin{equation}
R_{AB} = {\sum_{i=1}^H Z_{Ai}/\kappa_i \over  \sum_{i=1}^H Z_{Bi}/\kappa_i } \;\; \in Q
\end{equation}
are always rational numbers. Then it follows that there is a set of integers $\hat m_A$ such that
\begin{equation}
\sum_{i=1}^H {Z_{Ai}\over\kappa_i} = {\hat m_A\over \bar \kappa_*},
\end{equation}
where the $\hat m_A$ are all relatively prime. (Note $\bar \kappa_*$ does not have to equal $\kappa_*$). This is actually a (slightly) weaker condition than $R_{ij} = \kappa_i/\kappa_j$ being rational, since it is only if $Z_{Ai}$ is of rank $H$ that one can derive $R_{ij} \in Q$ from $R_{AB}\in Q$. Assuming  $R_{AB}\in Q$ the QNF master equation becomes
\begin{equation}
\sum_{A=1}^N   C_A \; \exp\left(  \hat m_A \; {\pi \omega\over \bar \kappa_*} \right) = 0.
\label{E:master2-1b}
\end{equation}
This can now be converted into a polynomial in exactly the same manner as previously, leading to families of QNFs as above. Provided both $R_{AB}$ and $R_{ij}$ are rational we can identify $\bar \kappa_* = \kappa_*/g$.

\section{Factorizability}\label{S:factor}

Now it is mathematically conceivable that in certain circumstances the master equation might factorize into a product over two disjoint sets of horizons
\begin{equation}
\left[\sum_{A=1}^{N_1}   C_{1A} \; \exp\left(  \sum_{i=1}^{H_1}  {Z_{1Ai} \; \pi \omega\over \kappa_{1i}} \right)\right]\; 
\left[\sum_{A=1}^{N_2}   C_{2A} \; \exp\left(  \sum_{i=1}^{H_2}  {Z_{2Ai} \; \pi \omega\over \kappa_{2i}} \right)\right] =  0.
\label{E:master3}
\end{equation}
Physically one might in fact expect this if the horizons indexed by $i\in\{1,\dots,H_1\}$ are very remote (in physical distance) from the horizons indexed by $i\in\{1,\dots,H_2\}$.  If such a factorization were to occur then the QNFs would fall into two completely disjoint classes,  being independently and disjointly determined by  these two classes of  horizon.   

While this might at first blush seem a physically plausible picture, mathematically however, this does not seem to happen (except in exceptional non-generic situations). For instance in the semi-analytic model reported in references~\cite{Skakala1, Skakala2}, (as  discussed above), this behaviour occurs only if $\cos(\pi\alpha_\pm)\to 0$.   Similarly, for master equations derived via monodromy techniques, the only physical situation in which this sort of factorization seems to occur is for spin 1 fields in a Schwarzschild-de Sitter (Kottler) background. So while the possibility of such a factorization at first looks plausible, it does not seem to be generic to the specific master equations of interest in this article.

\section{From ``families'' to constraints}\label{S:f2constraints}
\def\gap{{\;\mathrm{gap}}}
\def\B{{\mathcal{B}}}

We now wish to work ``backwards'' to see if the existence of a family of equi-spaced QNFs can lead to constraints on the ratios $R_{ij} = \kappa_i/\kappa_j$. Such an analysis has already been performed for the specific class of QNF master equations arising from semi-analytic techniques, and we now intend to generalize the argument to the generic class of QNF master equations presented in equation (\ref{E:master}). 
Let us therefore assume the existence of at least one ``family'' of QNFs of the form:
\begin{equation}
\omega_{n} = \omega_0 + in \gap; \qquad n\in\{0,1,2,3,\dots\}.
\label{E:family2}
\end{equation}
Then we are asserting
\begin{equation}
\sum_{A=1}^N   C_A \; \exp\left(  \sum_{i=1}^H  {Z_{Ai} \; \pi (\omega_0  + in \gap) \over \kappa_i} \right) = 0;  \qquad n\in\{0,1,2,3,\dots\}.
\label{E:master4}
\end{equation}
That is
\begin{equation}
\sum_{A=1}^N   \left\{ C_A \; \exp\left(  \sum_{i=1}^H  {Z_{Ai} \; \pi \omega_0 \over \kappa_i} \right) \right\} 
\exp\left(  i n \pi \gap \sum_{i=1}^H  {Z_{Ai} \over \kappa_i} \right)= 0;  \qquad n\in\{0,1,2,3,\dots\}.
\label{E:master4-2}
\end{equation}
We can rewrite this as
\begin{equation}
\sum_{A=1}^N  D_A 
\exp\left(  2\pi i n J_A \right)= 0;  \qquad n\in\{0,1,2,3,\dots\}.
\label{E:xxx}
\end{equation}
A priori, there is no particular reason to expect either the $D_A$ or the $J_A$ to be real. 

\subsection{Case 1}
One specific solution to the above collection of constraints is
\begin{equation}
\sum_{A=1}^N  D_A  = 0;  \qquad  \exp(2\pi i J_A)  =  r.
\label{E:master4-sp1}
\end{equation}
Furthermore, \emph{as long as no proper subset of the $D_A$'s sums to zero}, we assert that this is the only solution. To see this let us define
\begin{equation}
\lambda_A =  \exp(2\pi i J_A);   \qquad M_{AB} = (\lambda_A)^{B-1}; \qquad A,B \in \{1,2,3,\dots,N\}.
\end{equation}
Then $M_{AB}$ is a square $N\times N$ Vandermonde matrix, and then equation (\ref{E:xxx}) implies 
\begin{equation}
\sum_{A=1}^N  D_A  M_{AB} = 0,
\end{equation}
whence $\det(M_{AB}) =0$. But from the known form of the Vandermonde determinant we have
\begin{equation}
\det(M_{AB}) = \prod_{A>B} (\lambda_A - \lambda_B) = 0,
\end{equation}
implying that at least two of the $\lambda_A$ are equal. Without loss of generality we can shuffle the $\lambda_A$'s so that the two which are guaranteed to be equal are $\lambda_1$ and $\lambda_2$. Then equation  (\ref{E:xxx}) implies 
\begin{equation}
(D_1+D_2) \lambda_1^{B-1} +    \sum_{A=3}^N  D_A   (\lambda_A)^{B-1} = 0;  \qquad B\in\{1,2,3,\dots, N-1\}.
\end{equation}
But by hypothesis $D_1+D_2\neq 0$, so this equation can be rewritten in terms of a non-trivial reduced $(N-1)\times(N-1)$ Vandermonde matrix, whose determinant must again be zero, so that two more of the $\lambda_A$'s must be equal.  Proceeding in this way one reduces the size of the Vandermonde matrix by unity at each step and finally has
\begin{equation}
\lambda_A = r,
\end{equation}
as asserted. We then see
\begin{equation}
J_A = -i \,{\ln(r)\over2\pi} + m_A; \qquad  m_A \in Z.
\end{equation}
Expressed directly in terms of the surface gravities this yields
\begin{equation}
\sum_{i=1}^H Z_{Ai} {\gap\over \kappa_i} = -i \,{\ln(r)\over\pi} + 2 m_{A}; \qquad  m_{A} \in Z.
\end{equation}
By assumption, we have asserted the existence of at least one solution to these constraint equations. (Otherwise the family we used to start this discussion would not exist.) We have seen that we can choose to present the master equation in such a manner that  $\sum_{A=1}^N Z_{Ai}=0$. But then
\begin{equation}
0  = -i \,{\ln(r)\over\pi} N + 2 \sum_{A=1}^N m_{A}; \qquad  m_{A} \in Z.
\end{equation}
This implies that 
\begin{equation}
 i \,{\ln(r)\over\pi} = 2 \; {\sum_{A=1}^N m_{A}\over N}  = q  \in Q.
\end{equation}
That is, there is a rational number $q$ such that
\begin{equation}
\sum_{i=1}^H Z_{Ai} {\gap\over \kappa_i} = q + 2 m_{A}; \qquad  q\in Q; \qquad m_{A} \in Z.
\end{equation}
This is already enough to imply that the ratios $R_{AB}$ are rational.
If in addition $Z_{Ai}$ is of rank $H$ then, (either using standard row-echelon reduction of the augmented matrix, or invoking the Moore--Penrose pseudo-inverse and noting that the Moore-Penrose pseudo-inverse of an integer valued matrix has rational elements), we see that for each horizon the ratio $(\mathrm{gap})/\kappa_i$ must be a rational number, and consequently the ratios $R_{ij} = \kappa_i/\kappa_j$ must all be rational numbers. 

That is: If we have a family of QNFs as described by equation (\ref{E:family}), and if no proper subset of the $D_A$'s sums to zero, then using the observed features of the matrices $Z_{Ai}$ we can deduce that  the ratios $R_{AB}$ must all be rational numbers. With an additional hypothesis regarding the rank of the matrix $Z_{Ai}$,  we have seen that the  $R_{ij} = \kappa_i/\kappa_j$ must all be rational numbers. 

\subsection{Case 2}
More generally, \emph{if some proper subset of the $D_A$'s sums to zero}, subdivide the $N$ terms $A\in\{1,2,3,\dots, N\}$ into a cover of disjoint irreducible proper subsets $\B_a$ such that 
\begin{equation}
\sum_{A\in\B_a}^N  D_A  = 0.
\label{E:master4-sp2}
\end{equation}
Then the solutions of equation (\ref{E:xxx}) are uniquely of the from
\begin{equation}
\exp(2\pi i J_{A\in \B_a})  = \lambda_{A\in\B_a} = r_a.
\label{E:zzz}
\end{equation}
It is trivial to see that under the stated conditions this is a solution of equation~(\ref{E:xxx}), the only technically difficult step is to verify that these are the only solutions.  One again proceeds by iteratively using the Vandermonde matrix $M_{AB} = (\lambda_A)^{B-1}$ and considering its determinant. Instead of showing that \emph{all} of the $\lambda_A$'s equal each other, we now at various stages of the reduction process use the condition $\sum_{A\in\B_a}  D_A  = 0$ to completely decouple the corresponding $\lambda_{A\in\B_a} = r_a$ from the remaining $\lambda_{A\not\in\B_a}$. Proceeding in this way we finally obtain equation (\ref{E:zzz}) as claimed.

We then see
\begin{equation}
J_{A\in\B_a} = -i \,{\ln(r_a)\over2\pi} + m_{A\in\B_a}; \qquad  m_{A\in B_a} \in Z.
\end{equation}
Expressed directly in terms of the surface gravities this yields
\begin{equation}
\sum_{i=1}^H Z_{A\in\B_a,i} \; {\gap\over \kappa_i} = -i \,{\ln(r_a)\over\pi} + 2 m_{A\in\B_a}; \qquad  m_{A\in B_a} \in Z.
\end{equation}
With the obvious notation of $a(A)$ denoting the index of  the particular disjoint set $\B_a$ that $A$ belongs to, we can write this as
\begin{equation}
\sum_{i=1}^H Z_{Ai} \; {\gap\over \kappa_i} = -i \,{\ln\{r_{a(A)}\}\over\pi} + 2 m_{A}; \qquad  m_{A} \in Z.
\end{equation} 
This result now is somewhat more subtle to analyze. Let $A$ and $B$ both belong to a particular set $\B_a$. Then
\begin{equation}
\sum_{i=1}^H \{Z_{Ai}- Z_{Bi}\} {\gap\over \kappa_i} =  + 2 \{ m_{A} - m_B\} ; \qquad  m_{A}, m_B \in Z; \qquad A,B \in\B_a.
\label{E:11}
\end{equation} 
That is
\begin{equation}
\gap = {2\{m_{A} - m_B\} \over  \displaystyle \sum_{i=1}^H {\{Z_{Ai}- Z_{Bi}\} \over\kappa_i}};  \qquad \gap \in R; \qquad A,B \in\B_a;
\end{equation} 
so we see that the gap is real. (Furthermore, the gap is seen to be a sort of ``integer-weighted harmonic average'' of the $\kappa_i$.) But reality then implies that $r_a = e^{i\phi_a}$ so that
\begin{equation}
\sum_{i=1}^H Z_{Ai} {\gap\over \kappa_i} = {\phi_{a(A)}\over\pi} + 2 m_{A}; \qquad  m_{A} \in Z.
\end{equation}
By using $\sum_A Z_{Ai}=0$ we see that
\begin{equation}
{\bar\phi\over\pi} = \sum_a {\phi_a |\B_a|\over N \pi} \in Q,
\end{equation}
so that
\begin{equation}
\sum_{i=1}^H Z_{Ai} {\gap\over \kappa_i} = {\phi_{a(A)}-\bar\phi \over\pi} + 2 (m_{A}-\bar m); \qquad  m_{A} \in Z.
\end{equation}
Unfortunately in the general case there is little more than can be said and one has to resort to special case-by-case analyses. One last point we can make is that even though in this situation the $R_{ij}=\kappa_i/\kappa_j$ are sometimes irrational we can make the weaker statement that
\begin{equation}
{ \displaystyle \sum_{i=1}^H {\{Z_{Ai}- Z_{Bi}\} \over\kappa_i} \over  \displaystyle \sum_{i=1}^H {\{Z_{Ci}- Z_{Di}\} \over\kappa_i} } \;\;\;\in Q; \qquad A,B,C,D \in\B_a;
\end{equation} 
That is, certain weighted averages of the surface gravities are guaranteed to be rational. 
If we wish to analyze whether rational ratios of  $R_{ij}=\kappa_i/\kappa_j$ are implied in
each of the particular cases of interest, we need to:
\begin{itemize}
\item[a)] Check if there exists some $\omega_{0}$ giving non-trivial subsets $\B_{a}$, leading to $\eqref{E:master4-sp2}$.
\item[b)] Analyze the sets of equations $\eqref{E:11}$ implied by such an $\omega_{0}$.
\end{itemize}
By proceeding in this way we are able to prove that periodicity of the QNFs implies rational
ratios for the surface gravities in the following physically interesting cases:
\begin{itemize}
\item[a)] For $j=2m$ in equation $\eqref{E:SdS}$.
\item[b)] For equation $\eqref{E:RN}$ when  $j\neq 2m+1$ and $\cos(\pi j)\neq -\frac{1}{2}$.
\item[c)] For equation $\eqref{E:RNdS}$ when $j=2m$.
\end{itemize}
The proofs are quite long and tedious, without adding a lot of new
understanding, so we have decided to omit them in this paper.

\section{Discussion}\label{S:discussion}

We have seen that semi-analytic and monodromy techniques, although they are very different in both technical detail and underlying philosophy, by and large agree on the general \emph{form} of the QNF master equation they generate --- see equation (\ref{E:master}). For this entire class of master equations it is a rigorous result that whenever the ratio of surface gravities is rational $R_{ij} = \kappa_i/\kappa_j \in Q$, then the master equation can be reduced to a sparse polynomial and the QNFs fall into equi-spaced families of the type given in equation (\ref{E:family}).  This also holds true under the slightly weaker hypothesis that $R_{AB}\in Q$.  The converse of this result is more subtle: If there is at least one equi-spaced family of QNFs of the type given in equation (\ref{E:family}), then generically the ratio of surface gravities will be rational  $R_{ij} = \kappa_i/\kappa_j \in Q$, in which case all other QNFs will fall into families of this form. There are however some non-generic exceptional situations, which are rarely encountered but should be kept in mind. 

Even when general analyses based on our abstract form of the master equation are not quite as definitive as one might wish, the specific analyses based on the specific forms encountered in the literature often yield sufficient information to assert the rationality of either $R_{AB}\in Q$ or $R_{ij} = \kappa_i/\kappa_j \in Q$.



\end{document}